# Comment on "Evidence for a small hole pocket in the Fermi surface of underdoped YBa2Cu3Oy; [arXiv:1409.2788v1]"

Lev P. Gor'kov[1,2] and Gregory B. Teitel'baum[3]

[1]NHMFL, Florida State University, 1800 East Paul Dirac Drive, Tallahassee Florida 32310, USA and
[2]L.D. Landau Institute for Theoretical Physics of the RAS, Chernogolovka 142432, Russia
[3]E.K. Zavoiskii Institute for Technical Physics of the RAS, Kazan 420029, Russia

N. Doiron-Leyraud *et al.* in [1] report on the experimental observation of small holes pockets in the energy spectrum of $YBa_2Cu_3O_y$ (YBCO) coming below the transition into the charge ordered (CO) phase as a result of a Fermi surface (FS) reconstruction. We comment that their results seem to be more consistent with a unidirectional charge density wave (CDW) vector direction of which alternates along the c-axis between neighboring $CuO_2$-planes ($\vec{Q} \to Q_x, Q_y$).

In the interesting experimental study [1] the authors detected slow quantum oscillations (QOs) in the CO phase of YBCO which they attributed to holes pockets. The authors infer then that at low temperatures the electronic spectrum consists of one "diamond"-shape electronic pocket (*e*-pocket) bordering the Fermi arcs (FAs) at the nodal points in the Brillouin zone (BZ) and two smaller hole pockets (*h*-pockets) near the 'bare' Fermi surface located as shown in Fig.1c [1] on the both sides of the arc. That *h*-pockets are inherent only in the CO state was demonstrated convincingly by the very fact that slow QOs in the c-axis magneto-resistance disappear at the increase of temperature towards to $T_{CO}$-the temperature of the CO transition (Fig. 4b [1]).

It is well known that the angle-resolved photoemission spectroscopy (ARPES) in several cuprates reveals coherent excitations only on some finite "arcs" along the 'bare' Fermi surface (FS) called the Fermi arcs (FAs) [2-4]. Although ARPES is not possible yet in every cuprate compound, the consensus is that the feature bears a general character and FAs are ubiquitous for all underdoped (UD) cuprates. Thereby carriers on the FAs constitute one component of the energy spectrum (holes) responsible, in particular, for the Fermi-liquid-like regime of the $T$-square resistivity at temperatures below $T^{**}(x) < T^*(x)$ [5, 6] (here $T^*(x)$ is the pseudogap (PG) temperature and $x$-the holes concentration).

The electronic pocket (*e*-pocket) first manifested itself in low temperature QOs [7] and since that the view prevailing in the literature is that *e*-pocket arises simultaneously with the FS reconstruction at some phase transition [7, 8]. Recent X-ray and NMR experiments disclosed in UD YBCO the tendency to CDW instabilities. The latter finally realize themselves in a charge order (CO) phase transition at typical temperatures $T_{CO} \sim (40 \div 50)K$ [9-13]. Attribution of the *e*-pocket to reconstruction of the FS at the CO transition seems however to encounter the difficulty that in YBCO and in the single-layer mercury compound Hg1201 electrons actually make themselves visible already above $T \sim 100$ K via the negative contributions into the temperature-dependent Hall coefficient (see e.g. in [14]). (More evidences suggesting that in reality the *e*-pocket exists already well above the CO transition were discussed recently in [15]).

We address the nature of both *e*-and *h*-pockets in frameworks of the FAs concept. We argue that appearance of new QOs [1] in the CO phase is more consistent with the uniaxial charge density wave (CDW) vector $\vec{Q}_o$ rotating between the neighboring $CuO_2$-planes by ninety degrees ($\vec{Q}_0 \to Q_{0,x}, Q_{0,y}$). We also find, among other things, that CDW instabilities are detrimental for carriers on FAs.

Choose for sake of argument the unidirectional CDW with the structural vector $\vec{Q}_0 = (Q_{0,x}; 0)$ (by its value $|Q_0|$ is of the order of few tenths of $2\pi/a$, where $a$ -the lattice period [7-11]). In Fig.1a the thick red arcs are all four FAs in the tetragonal BZ. The two arcs on the left are also shown being shifted to the right by the vector CDW $\vec{Q}_0 = (Q_{0,x}; 0)$. Points ***A, A′*** are the ending points for the upper two arcs in the BZ (correspondingly, ***A, A′*** are the ending points of the two arcs from below); the dashed lines show the locus of the 'bare' FS "buried" under the pseudo gaps along the antinodal directions.

To be specific, introduce $M(\vec{p}; p_x \pm Q_x) = U_0 n(q) \exp(\pm i Q_x)$ - the matrix element for scattering of holes on FAs with a momentum $\vec{p}$ on the CDW fluctuations; here the Fourier component of density $n(q) \equiv n(q \pm Q_{0,x})$ is in the dimensionless units: $n(x) \to n(x)/a^2$; $U_0$ is a typical energy scale for cuprates ($\sim 1 eV$). (For shortness, the sub-index in $Q_{0,x}$ will be omitted below). Let also $\varepsilon(\vec{p})$ be any phenomenological band spectrum from [16-19]. In the *fluctuation* regime $T_{onset} > T > T_{CO}$:

$$n(x) = \Sigma_{\pm} \int n(q) dq \exp[\pm i(Q_0 + q)(x - x_0)]. \quad (1)$$

The interaction $M(\vec{p}; p_x \pm Q_x)$ does not affect directly the spectrum of carriers on the Fermi arc, but transitions between two states with the two momentums $\vec{p}, \vec{p}'$ *at the arc* become possible in the second order of the ordinary quantum mechanical perturbation theory. Introducing the notation $W(p_x, p'_x; p_y) = <\Sigma_{\pm,l} \hat{M}(\vec{p}; \vec{l} \pm Q)\hat{M}(\vec{l} \mp Q; \vec{p}')>$ for the latter obtain:

$$W(p_x, p'_x; p_y) = |U_0|^2 \ e^{i(p'_x - p_x)x_0} \int \frac{dl}{2\pi} \times \Sigma_{\pm} \left[ \frac{n(p_x - l)n(l - p'_x)}{\varepsilon(p_x; p_y) + \mu - \varepsilon(p_x + l_x \pm Q_0; p_y)} \right]. \quad (2)$$

Following [7], for the convenience of calculations the fluctuations $n(q)$ are taken in the Gaussian form:

$$n(q) = (\bar{n}(T)/\sqrt{2\pi})\xi \exp[-(q\xi)^2/2]. \quad (3)$$

(At $\xi \to \infty$ the expression transforms into the $\delta$-function: $(1/\sqrt{2\pi})\xi \exp[-(q\xi)^2/2] \Rightarrow \delta(q)$; correspondingly, below $T_{co}$ $\bar{n}(T)$ in Eq. (3) goes over in the CO order parameter: $\bar{n}(T) \to \bar{n}_0(T)$).

If the coherence length $\xi(T)$ for CDW fluctuations is large $\xi \gg a$ in $\varepsilon(p_x + l_x \pm Q_0; p_y)$ one can omit the dependence on $l_x$. In (2) $p_x, p_y$ are in a vicinity of the arc; so that the energy $\varepsilon(p_x; p_y) \equiv \varepsilon(\vec{p})$ (counted from the chemical potential $\varepsilon_F \equiv \mu$) is small; large is the value of $|\mu - \varepsilon(p_x + l_x \pm Q_0; p_y)|$. Substituting $n(q)$ from (3) after integrating over $l_x$ one finds:

$$W(p_x, p'_x; p_y) = \frac{|U_0|^2}{\bar{E}(p_x \pm Q; p_y)} \left(\frac{\bar{n}(T)^2 \xi}{4\pi^{3/2}}\right) e^{i(p'_x - p_x)x_0} \exp\left[-\frac{(p_x - p'_x)^2 \xi^2}{4}\right]. \quad (4)$$

(Here $\bar{E}(\pm Q; p_y) = |\mu - \varepsilon(p_x \pm Q_0; p_y)|$ stands for the characteristic energy in (2). Notice the dependence on the position $p_y$ along the arc).

One could proceed further and calculate the inverse scattering time $1/\tau_{FA,CDW}$ for scattering of the arc's carriers on fluctuations in the incommensurate (IC) CDW:

$$\frac{1}{\tau_{FA,CDW}} = (\pi^{3/2}/\sqrt{2})\bar{n}(T)^4 \left(\frac{\xi(T)p_F}{\Delta\varphi}\right)\left\{\frac{1}{2}\sum_{\pm}\left[\frac{|U_0|^2}{\varepsilon_F \bar{E}(p_x \pm Q; p_y)}\right]\right\}^2. \quad (5)$$

The order parameter of the CO phase gradually grows below $T_{co}$ on the background of continuing fluctuations. Substitution of $n_{CO}(x) = \bar{n}_0(T)\exp[\pm iQ_0(x-x_0)]$ into Eq. (2) ($p_x \equiv p'_x$) gives the following correction to the energy spectrum on FAs:

$$\Delta\varepsilon(\vec{p}) = \sum_{\pm} < M(\vec{p}; \vec{p} \pm Q_0) \frac{1}{\varepsilon(\vec{p}) + \mu - \varepsilon(\vec{p} \pm Q_0)} M(\vec{p} \mp Q_0; \vec{p}) >. \quad (6)$$

Points **B** and **<u>B</u>** in Fig.1a correspond to zeros in the denominator in (6): $\varepsilon(\vec{p}) = \mu - \varepsilon(\vec{p} \pm Q_0)$ and are two singular points at which two branches of the spectrum (those on the right arcs and on the arcs from the left) become split in the presence of CDW of non-zero amplitude; the resulting energy gaps are shown in Fig. 1b. (The expressions for the spectrum near two points **B** and **<u>B</u>** are not written down here because most parameters in Eq. (6) are unknown experimentally. By the order of magnitude, an estimate for the gap follows from the value of the Fermi energy for the holes pocket $E_{F,h} \approx (200 \div 300)K$ [1]).

When the momentum $\vec{p}$ is not too close to **B** and **<u>B</u>**, $\Delta\varepsilon(\vec{p})$ signifies a correction to the arc's spectrum in the CO phase. As to the holes pockets, they can form only inside a triangular area in Fig.1b between points **B** (**<u>B</u>**) and **A, A'** (**<u>A</u>, <u>A'</u>**) on the 'bare' FS. The corresponding energy potential for that is composed of the new gapped spectrum close to **B** (**<u>B</u>**) and large antinodal pseudogaps. (The elliptic shape of *h*- pockets in Fig.1b serves only for illustration). For the *h*-pockets to exist the lengths of the Fermi arcs in Fig.1b should be large enough. Note in passing, in that connection, that the length of FAs increases with the increase of dopants' concentration [15].

Notice that a uniaxial CO parameter with directions of the CDW vector alternating along the c-axis between the two $CuO_2$-planes in YBCO $\vec{Q} \to Q_x, Q_y$ would simulate the pattern of the two *h*-pockets *on the both sides* on the single arc shown in Fig.1c [1].

The problem hence is about origin and location of the electronic pocket. In case of the biaxial CO parameter of large enough amplitude the mathematical construction leading to the energy spectrum of the form shown in Fig.1c [1] could be, in principle, no contradictory.

The very question whether the CDW structural vector in YBCO is uniaxial or biaxial is not resolved experimentally [10, 13, 20]. It is also worth emphasizing that the experiments on QOs leave undefined the positions of all pockets in the BZ.

In this respect, we point out on several observations of an experimental character that seem to be incompatible with positioning of the electronic pocket as it is shown in Fig.1c [1].

In the first place, an *e*-pocket bordering the 'bare' FS at the nodal point will be gapped at Cooper pairing. Meanwhile, experimentally, in YBCO electrons on the pocket remain in the normal state even in the *superconducting* phase revealing in the specific heat linear-in-temperature term down to 5K [21]. That is possible [22] only if *e*-pocket were centered at the $\Gamma$–point of the Brillouin zone of the approximately tetragonal symmetry.

Secondly, as it was mentioned in the beginning, the negative contributions into the Hall coefficient from the electronic pocket are seen experimentally at temperatures ~100 K, i. e., considerably higher than $T_{CO} \sim (40 \div 50)K$ [14]. The Hall coefficient continues to decrease in the CO phase without discontinuity at $T_{CO}$ (in [14] superconductivity is suppressed by high magnetic fields).

Finally, all excitations scatter on CDW fluctuations. Thanks to the smaller *e*-pocket's size $p_{Fe} \approx (1/5 \div 1/6)p_F$ scattering of electrons on fluctuation of IC CDW expected to be weaker as electrons feel the averaged potential of a short-wavelength CDW. Fortunately, the mobility of electrons in the CO phase can be found from the value of the Dingle temperature $T_D \approx 6K$ [21] and hence is already temperature independent at such temperatures. The inverse relaxation time for the FAs excitations is given by Eq. (5). With all energies in (5) being of the same order $U_0$ ($\sim 1eV$), $\bar{n}(T) \sim 1/10$, $\xi(T)p_F \sim 0.01$ and $\Delta\varphi \approx 0.1$ [15] $1/\tau_{FA,CDW}$ can be as large as $\sim 0.1 eV$ thereby smearing away the coherence of excitations on the Fermi arcs at low temperatures.

In summary, we thus arrive to the conclusion that the uniaxial CO parameter in combination with an electronic pocket at the $\Gamma$-point of the BZ preexisting at the CO transition suggests a more plausible interpretation of the low frequency QOs discovered in [1].

**Acknowledgements.** The work of L. P. G. was supported by the NHMFL through NSF Grant No. DMR-1157490, the State of Florida and the U.S. Department of Energy; that of G. B. T. by the Russian Academy of Sciences through Grants No. P 20 and No. OFN 03.

### References

[1] N. Doiron-Leyraud et al., arXiv:1409.2788v1 (2014).

[2] Yoshida et al., J. Phys. Soc. Jpn. **81**, 011006 (2012).

[3] Yoshida, T. et al. Phys. Rev. B **74**, 224510 (2006)

[4] D. Fournier et al., Nature Physics **6**, 905 (2010).

[5] Barišić, N. et al. Proc. Natl. Acad. Sci. **110**,12235 (2013).

[6] Gor'kov, L. P., Phys. Rev. (Rapid Comm.) B **88**, 041104 (2013).

[7] Doiron-Leyraud, N. et al. Nature **447**, 565-568 (2007).

[8] Taillefer, L. J. Phys.: Condens. Matter **21** 164212 (2009).

[9] Chang, J. et al. Nature Phys. **8**, 871-876 (2012).

[10] Ghiringhelli, G. et al., Science **337**, 821-825 (2012).

[11] Blackburn, E. et al. Phys. Rev. B **88**, 054506 (2013).

[12] Blackburn, E. et al. Phys. Rev. Lett. **110**, 137004 (2013).

[13] Wu, T. et al. Nature **477**, 191-194 (2011).

[14] Doiron-Leyraud, N. et al. Phys. Rev. X **3**, 021019 (2013).

[15] L.P. Gor'kov and Gregory B. Teitel'baum, arXiv:1407.5888v2 (2014).

[16] A. J. Millis and M. R. Norman, Phys. Rev. B **76**, 220503(R), (2007).

[17] J.-X. Li, C.-Q Wu and D.-H. Lee, Phys. Rev. B **74**, 184515 (2006).

[18] K. Seo, H.-D. Chen and J. Hu, Phys. Rev. B **76**, 020511(R) (2007).

[19] A. Allais, D. Chowdhury and S. Sachdev, arXiv:1406.0503 (2014).

[20] R. Comin et al., arXiv:1402.5415.

[21] S. C. Riggs et al., Nature Phys. **7**, 332-335 (2011).

[22] L. P. Gor'kov, Phys. Rev. B **86**, 060501 (2012).

**Captions to the figures**

**Figure1. Formation of the holes pocket at onset of the charge ordering: a**) Four FAs in the tetragonal BZ are shown by the thick red lines. The left arcs are also drawn being shifted to the right at interaction with a CDW potential with the unaxial vector $\vec{Q} = (Q_x; 0)$. Points *A, A'* and *A, A'* are the ending points of the arcs; the dashed lines show locus of the 'bare' FS "buried" under the pseudo gaps seen in ARPES along the antinodal direction. **b**) Eq. (6) gives the contribution into the renormalized energy spectrum away from points *B* and *B*. The two points corresponding to crossing of the FAs in Fig.1a are singular in that in their vicinity the two branches of the initial spectrum become split by the CO transition. Below temperature of the CO transition the spectrum near the two points becomes gapped and the holes pockets (pink color) may form in the two areas of a triangular shape each via the potentials owed to the new gaps at the *B*, *B* points and that of the large pseudogaps along the 'bare' FS at the points *A, A'* and *A, A'*. (Elliptic shape for the *h*-pocket is chosen for illustration only). Scattering of the FA excitations on the CDW fluctuations is strong smearing this branch of the energy spectrum at low temperatures (see text).

Not to overcrowd the figures the electronic pocket at the Γ-point is not shown.

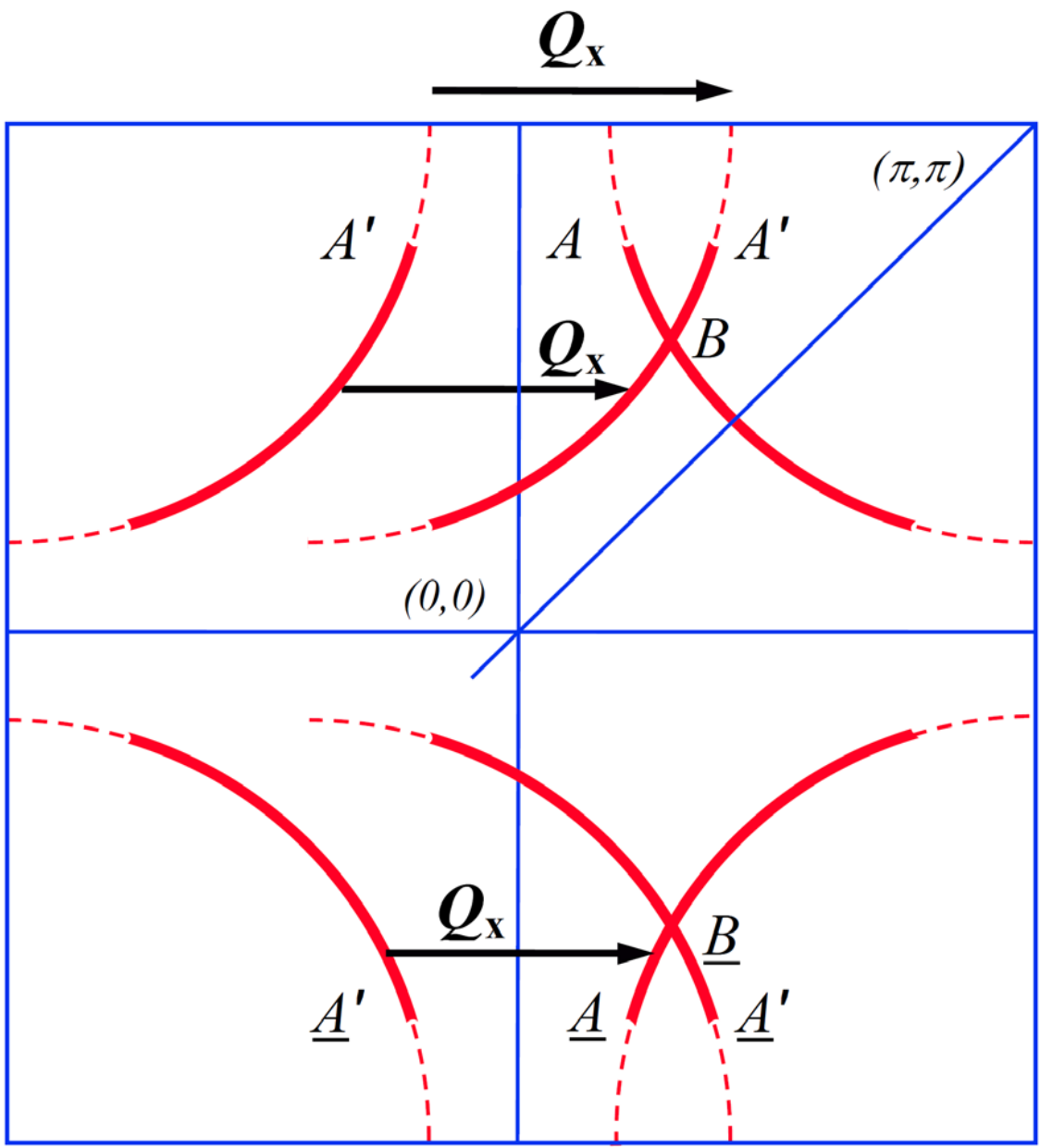

Fig. 1a.

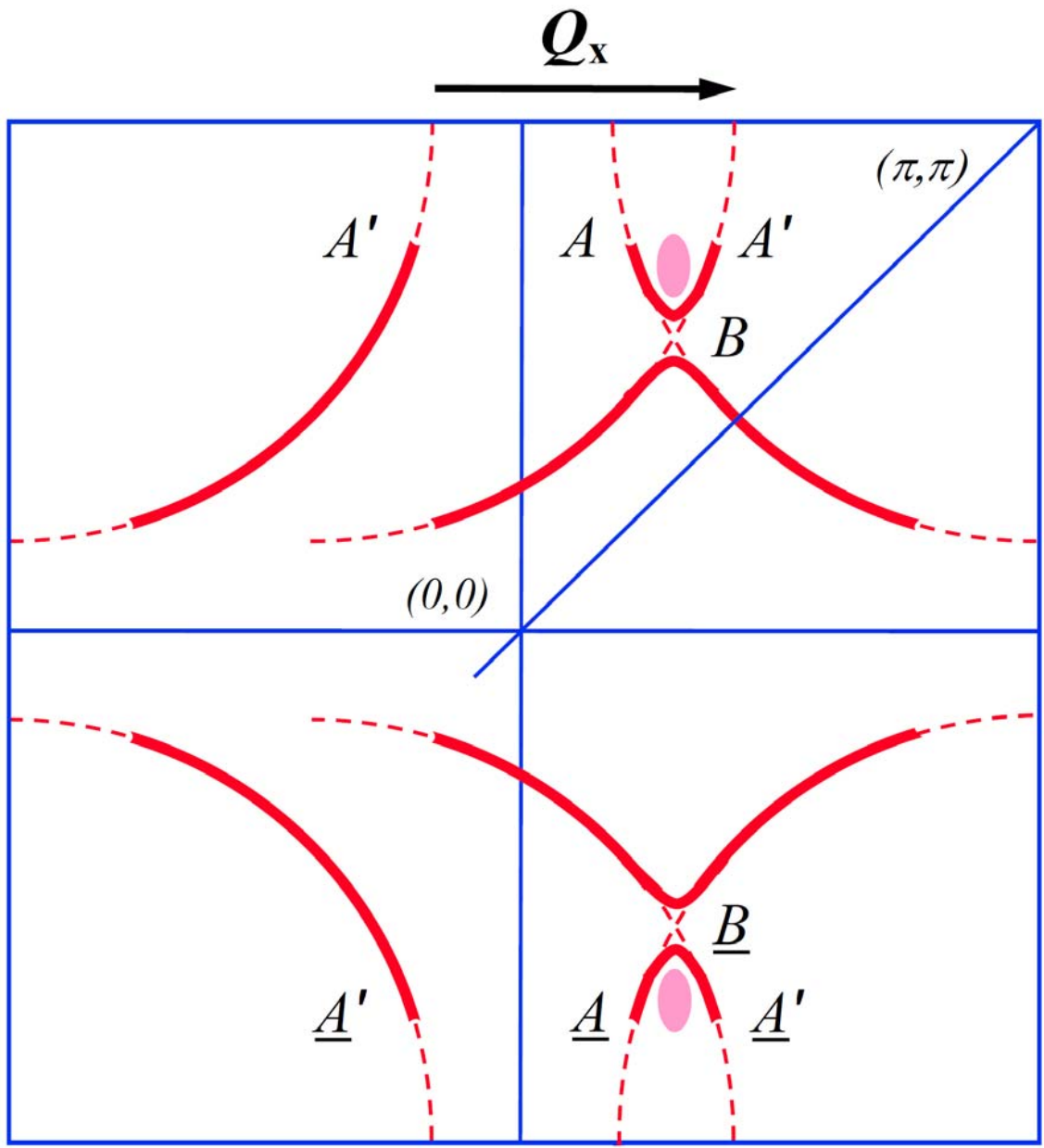

Fig. 1b.